\documentclass[twocolumn]{aastex7}

\usepackage{hyperref}
\usepackage{amsmath}
\usepackage{gensymb}
\usepackage{booktabs}
\usepackage{bm}
\usepackage{tablefootnote}
\usepackage{natbib}
\usepackage{CJK}
\usepackage{xspace}
\usepackage{enumitem}

\newcommand{\Atwentyfive}{\hyperlink{cite.attia2025JADE}{A25}\xspace}

\begin{document}
\begin{CJK*}{UTF8}{gbsn}
\title{Observational and Dynamical Constraints on an Unseen Outer Perturber in the GJ 436 Hot Neptune System}
\author[0009-0006-0871-1618]{Haedam Im (\CJKfamily{mj}임해담)}
\altaffiliation{Dorrit Hoffleit Undergraduate Research Fellow}
\affiliation{Department of Physics, Massachusetts Institute of Technology, Cambridge, MA 02139, USA
}
\affiliation{Department of Astronomy, Yale University, 219 Prospect Street, New Haven, CT 06511, USA}
\email{haedamim@mit.edu}

\author[0000-0003-0834-8645]{Tiger Lu (陆均)}
\altaffiliation{Flatiron Research Fellow}
\affiliation{Center for Computational Astrophysics, Flatiron Institute, 162 5th Avenue, New York, NY 10010, USA}
\affiliation{Department of Astronomy, Yale University, 219 Prospect Street, New Haven, CT 06511, USA}

\author[0000-0002-7670-670X]{Malena Rice} 
\affiliation{Department of Astronomy, Yale University, 219 Prospect Street, New Haven, CT 06511, USA}

\author[0000-0001-6532-6755]{Quang H. Tran} 
\altaffiliation{51 Pegasi b Fellow}
\affiliation{Department of Astronomy, Yale University, 219 Prospect Street, New Haven, CT 06511, USA}

\author[0000-0001-8308-0808]{Gongjie Li}
\affiliation{Center for Relativistic Astrophysics, School of Physics, Georgia Institute of Technology, Atlanta GA 30332, USA}

\author[0000-0002-9802-9279]{Smadar Naoz}
\affiliation{Department of Physics and Astronomy, UCLA, Los Angeles, CA 90095, USA}
\affiliation{Mani L. Bhaumik Institute for Theoretical Physics, Department of Physics and Astronomy, UCLA, Los Angeles, CA 90095, USA}

\begin{abstract}
Hot Neptunes in the sub-Jovian desert offer unique insights into planetary system evolution, retaining signatures of dynamical processes that shaped their present-day architectures. Many of these planets exhibit polar orbits, yet the mechanisms responsible for these misalignments between the stellar spin axis and planet orbit normal remain under debate. GJ 436 b stands among the very few hot Neptunes with both a polar and an eccentric orbit, thereby preserving dynamical signatures that may have otherwise been erased by tidal circularization. We investigate the unusual orbital architecture of GJ 436, exploring von Zeipel-Lidov-Kozai migration induced by a distant companion as a mechanism to explain the present-day orbit of GJ 436 b. Using $\sim$20~years of archival radial velocity measurements and astrometric data from the Hipparcos-Gaia Catalog of Accelerations, we constrain a potential companion to $a_{\mathrm{c}}<5.4$~AU for $m_{\mathrm{c}}>0.05~M_{\mathrm{Jup}}$ and $a_{\mathrm{c}}<64$~AU for $m_{\mathrm{c}}>24~ M_{\mathrm{Jup}}$ in the GJ 436 system at the $2\sigma$ confidence level, providing the most stringent constraints to date. We further perform three-body hierarchical secular simulations to determine which companion configurations could reproduce GJ 436 b's present-day orbit within the observationally constrained parameter space. Our dynamical modeling favors sub-Jovian masses on orbits with $a_\mathrm{c} \gtrsim 6.8$ AU, suggesting a substellar perturber. These observational and dynamical constraints can guide future companion searches and illuminate formation mechanisms for hot Neptune desert planets on polar orbits. 
\end{abstract}

\section{Introduction \label{introduction}}

A small but growing population of Neptune-sized planets on close-in orbits---commonly referred to as ``hot Neptunes''---offers intriguing insights into planetary formation and dynamical processes. 
The ``sub-Jovian desert" or ``Neptune desert" refers to the scarcity of such hot Neptunes, with radii $2~R_\oplus\lesssim R_p \lesssim10~R_\oplus$ and orbital periods $P\lesssim3$ days \citep{SzaboKiss2011,LundkvistKjeldsen2016, MazehHolczer2016, hallattmillholland2025desert}. Long thought to be empty, in recent years missions such as the Transiting Exoplanet Survey Satellite \citep[TESS;][]{RickerWinn2015} have begun to populate this region with discoveries. Today, over twenty confirmed planets lie within the Neptune desert \citep[e.g.][]{JenkinsDiaz2020, ArmstrongLopez2020, PerssonGeorgieva2022, NaponielloMancini2023, OsbornArmstrong2023}. 

Interestingly, multiple planets in and near the Neptune desert are significantly spin-orbit misaligned (that is, the planet's orbit normal axis is significantly offset from its host star's rotational axis), including GJ-3470 b \citep{StefanssonMahadevan2022}, TOI-3884 b \citep{Libby-RobertsSchutte2023}, and WASP-107 b \citep{DaiWinn2017, RubenzahlDai2021}. The nature of this so-called preponderance of perpendicular planets \citep{albrecht_perponderance} is not fully understood \citep{siegel_ponderings,Dong23}, but is nonetheless puzzling because planets are typically expected to form in circular, aligned orbits due to strong eccentricity and inclination damping from the protoplanetary disk \citep[e.g.][]{cresswell2007evolution}.

Several mechanisms have been proposed to explain the inclined orbits of close-in planets (including, but not limited to, hot Neptunes). For example, strong planet-planet scattering events are capable of exciting spin-orbit misalignments \citep[e.g.,][]{chatterjee_2008, nagasawa2008scattering, nagasawa2011orbital}, and in systems with 3 or more bodies, von Zeipel-Lidov-Kozai \citep[ZLK;][]{von_zeipel_1910, lidov1962evolution, kozai1962secular,naoz2016eccentric} oscillations can generate eccentric and polar orbits \citep[e.g.][]{wu_2003,naoz_2011,vick_misaligned_zlk,LuAn2025}. Disk-driven secular resonances can also produce large inclinations during protoplanetary disk dissipation \citep{petrovich_2020}. Finally, an array of mechanisms can generate large primordial misalignments by tilting either the stellar spin axis or the protoplanetary disk \citep[e.g.][]{bouvier1999magnetospheric, lai2008wave, LaiFoucart2011, ginski2021disk, kuffmeier2021misaligned}.

These polar Neptunes typically have circular orbits, suggesting that on a population level, the ages of these systems ($t_{\mathrm{age}}$) are younger than their spin-orbit realignment times ($t_{\mathrm{align}}$) but older than their circularization times ($t_{\mathrm{circ}}$), or $t_{\mathrm{circ}} < t_{\mathrm{age}}<t_{\mathrm{align}}$. However, there is one exception to this trend---GJ 436 b, a Neptune-sized planet on a close-in, eccentric and polar orbit \citep{ButlerVogt2004,GillonPont2007}. GJ 436 b has become one of the most extensively studied exoplanets over the past two decades, through a series of both atmospheric characterization and precise transit and radial velocity (RV) measurements (e.g., \citealt{KnutsonFulton2014, LavieEhrenreich2017, LothringerBenneke2018, Guzman-MesaKitzmann2022, RosenthalFulton2021,KokoriTsiaras2023, MukherjeeSchlawin2025, FinnertyFitzgerald2026}). 

Multiple studies have consistently measured a substantial orbital eccentricity ($e \gtrsim 0.12$; e.g., \citealt{ButlerWright2006, ManessMarcy2007, WrightMarcy2007, KnutsonFulton2014}) for GJ 436 b. Additionally, \cite{BourrierLovis2018} discovered a large spin-orbit misalignment via the Rossiter--McLaughlin effect \citep{rossiter1924detection, mcLaughlin1924some}, later refined to $\Psi_{\mathrm{Ab}} = {103.2}_{-11.5}^{+12.6}{}\degree$ \citep{BourrierZapateroOsorio2022}. There are three potential solutions to this tension: either (1) the aforementioned timescale ordering does not hold, (2) the planet was delivered to its current orbit relatively recently, or (3) an unusual dynamical mechanism is responsible for maintaining GJ 436 b's eccentricity. The planet is therefore a particularly valuable probe of the mechanisms responsible for generating polar Neptunes: its preserved orbital eccentricity signals a dynamically hot history and serves as a signature of earlier evolution that would have been erased in other, fully circularized systems.

The orbit of GJ 436 b has hence been the subject of great interest since its discovery, and numerous evolutionary pathways have been set forth to explain the system configuration \citep[e.g.][]{RibasFont-Ribera2008, batygin2009gj436, beust_2012}. These scenarios invariably demand the presence of an external companion in the system. However, to date no such perturber has been definitively discovered. 

Several observational studies have attempted to detect this putative companion. \cite{RibasFont-Ribera2008} initially claimed the detection of a $\sim$5 M$_\oplus$ super-Earth with a 5.2-day orbital period in a 2:1 mean-motion resonance based on transit timing variations (TTVs), but this was refuted by \cite{2008A&A...487L...5A}. Later, \cite{BallardCharbonneau2010} investigated a potential 0.75~$M_\oplus$ candidate planet from Extrasolar Planet Observation and Deep Impact Extended Investigation \citep[EPOXI;][]{KlaasenA'Hearn2013} data, but subsequent Spitzer Space Telescope \citep{WernerRoellig2004} observations revealed the signal to arise from instrumental noise. \cite{StevensonBean2014} proposed two sub-Earth candidates (UCF-1.01, UCF-1.02) via Spitzer observations, but subsequent Hubble Space Telescope (HST) observations yielded non-detections. Furthermore, \cite{MaciejewskiNiedzielski2014} found no evidence of TTVs in this system based on 3 years of photometric data with the 2.2-m telescope at the Calar Alto Observatory \citep{SanchezAceituno2007} and the 2.56-m Nordic Optical Telescope \citep{BaldwinTubbs2001}. Similarly, \cite{LanotteGillon2014} concluded that a joint analysis of HARPS \citep{PepeMayor2000} and Spitzer data does not find evidence for an outer planet. Despite many attempts, the putative GJ 436 c remains elusive.

In this paper, we present the most extensive analysis of the GJ 436 system to date, using a $\sim$20-year baseline of archival RV measurements combined with astrometric data from the Hipparcos-Gaia Catalog of Accelerations \citep[HGCA;][]{brandt2018hgca,brandt2021hgca}. We complement these observational constraints with three-body secular dynamical analyses and simulations to further predict and constrain the properties of a potential outer perturber in the GJ 436 system. In \S\ref{Observations}, we describe our joint analysis of RV and astrometric observations. We do not detect an additional planet, but we place stringent constraints on the mass and orbital parameters of a hidden companion should it exist. In \S\ref{theoretical background}, we leverage simple analytic arguments in the case of ZLK migration to dynamically constrain the parameter space of a potential GJ 436 c. In \S\ref{dynamical_sims}, we use a secular code to simulate ZLK migration of GJ 436 b, further dynamically constraining the allowable parameter space. In \S\ref{attia}, we discuss methodological differences between this work and the recent similar study of \cite{attia2025JADE}, hereafter \Atwentyfive. In \S\ref{Conclusions}, we present our conclusions. 

\begin{deluxetable}{lcc}[t]
\tablecaption{Table of RV fit priors and best-fit values }
\tablewidth{\columnwidth}
\tablehead{ 
\colhead{Parameter} & \colhead{Priors} & \colhead{Best-fit values}}
\startdata
$P_b$ (days) & $\mathcal{N}(2.643892, 2.5\times10^{-5})$ & $2.6438904^{+8.3\times10^{-6}}_{-7.5\times10^{-6}}$ \\
$T_{\text{conj},b}$ (JD) & $\mathcal{N}(2454428.848, 0.285)$ & $2454428.860^{+0.008}_{-0.013}$ \\
$e_{\mathrm{b}}$ & $\mathcal{U}(0, 0.99)$ & $0.155^{+0.0123}_{-0.0084}$ \\
$\omega_{\mathrm{b}}$ (rad) & --- & $-0.710^{+0.087}_{-0.054}$ \\
$K_{b}$ (m s$^{-1}$) & --- & $17.42^{+0.14}_{-0.27}$ \\
$m_{\mathrm{b}}\sin i~(M_{\oplus})$ & --- & $21.57^{+0.33}_{-0.47}$ \\
$\gamma_{\textrm{\scriptsize HIRES1}}$ & --- & $-2.09^{+0.75}_{-0.62}$ \\
$\gamma_{\textrm{\scriptsize HIRES2}}$ & --- & $-0.83^{+0.69}_{-0.62}$ \\
$\gamma_{\textrm{\scriptsize CARMENES}}$ & --- & $-19.5^{+1.5}_{-1.3}$ \\
$\gamma_{\textrm{\scriptsize HARPS}}$ & --- & $9791.31^{+0.61}_{-0.49}$ \\
$\dot{\gamma}$ & --- & $-0.0002^{+0.00058}_{-0.00063}$ \\
$\ddot{\gamma}$ & --- & $-5\times10^{-8}{}^{+1.65\times10^{-7}}_{-1.45\times10^{-7}}$ \\
$\sigma_{\textrm{\scriptsize HIRES1}}$ & $\mathcal{U}(0, 1000)$ & $3.0^{+1.1}_{-0.3}$ \\
$\sigma_{\textrm{\scriptsize HIRES2}}$ & $\mathcal{U}(0, 1000)$ & $3.64^{+0.44}_{-0.13}$ \\
$\sigma_{\textrm{\scriptsize CARMENES}}$ & $\mathcal{U}(0, 1000)$ & $1.72^{+0.53}_{-0.14}$ \\
$\sigma_{\textrm{\scriptsize HARPS}}$ & $\mathcal{U}(0, 1000)$ & $1.17^{+0.30}_{-0.08}$
\enddata

\tablecomments{Jitter terms $\sigma$ and RV zero points $\gamma$ are in m s$^{-1}$. The linear RV trend $\dot{\gamma}$ is in m s$^{-1}$ d$^{-1}$, and the quadratic RV acceleration $\ddot{\gamma}$ is in m s$^{-1}$ d$^{-2}$. The reference epoch for $\gamma$, $\dot{\gamma}$ and $\ddot{\gamma}$ is 
JD 2453422.3635. HIRES1 and HIRES2 correspond to the Keck/HIRES instrument before and after its 2004 upgrade, respectively. The planetary mass $m_{\mathrm{b}}$ is calculated using the stellar mass value reported by \cite{RosenthalFulton2021}.}
\label{tab: radvel table}
\end{deluxetable}

\section{Observations and Data Analysis \label{Observations}}
\subsection{Radial Velocity Fit}

\begin{figure*}
    \centering
    \includegraphics[width=\textwidth]{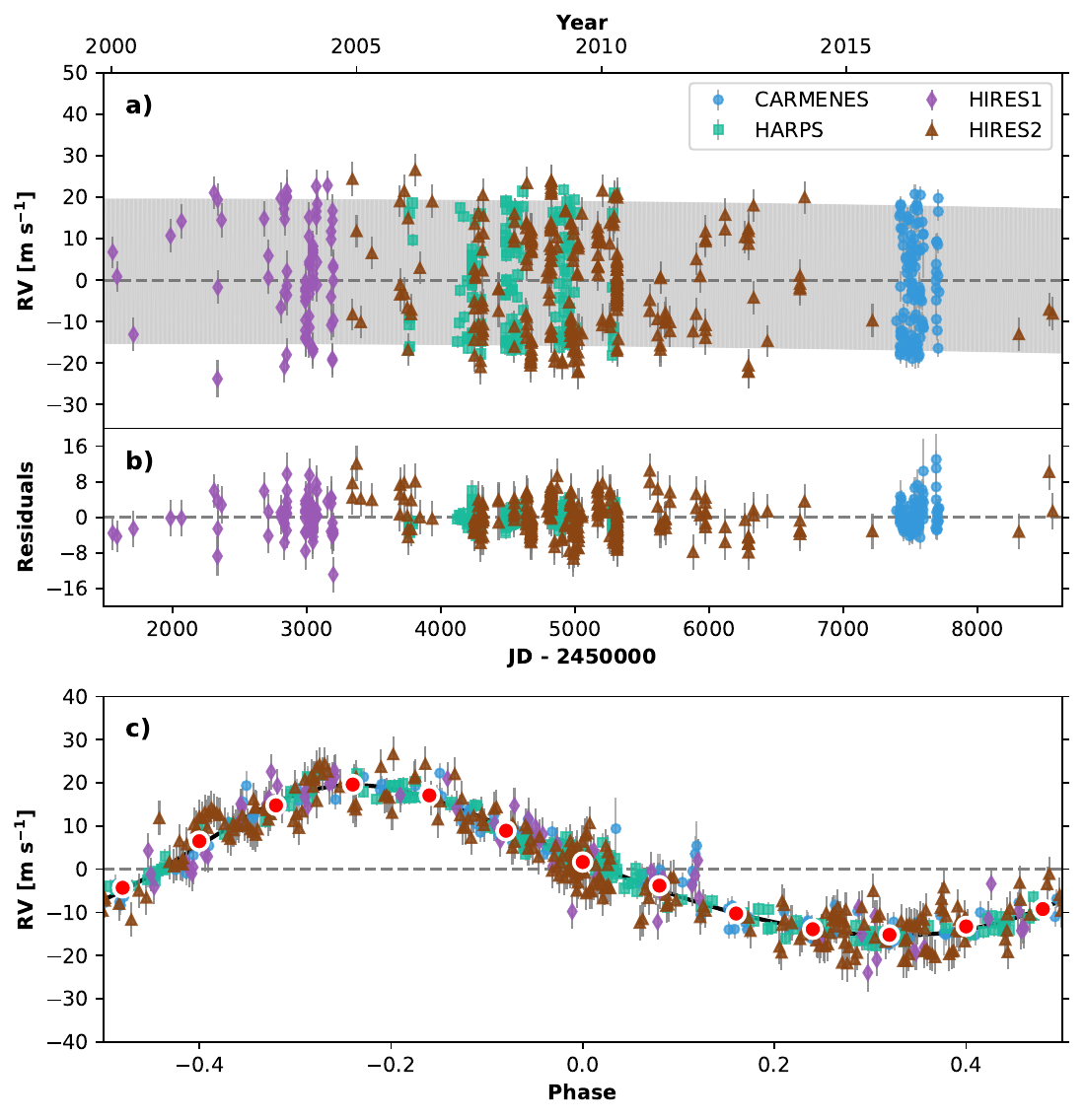}
    \caption{RV measurements and our best-fitting model of the GJ 436 system. Panel (a) shows the RV data with the best-fit model  over the 19-year baseline. Panel (b) displays the RV residuals. Panel (c) presents the phased data and model based on the best-fit orbital parameters. Each instrument is plotted with a different color, and the Keplerian model is shown in black.}
    \label{fig:RV_plot}
\end{figure*}

We compile 629 RV measurements spanning 19 years from three independent studies. We use 345 RV measurements from the California Legacy Survey \citep[CLS;][]{RosenthalFulton2021} spanning 19.2 years from 2000 to 2019 with the HIRES spectrograph ($\lambda$ = 380--690~nm, $R = $~55,000--86,000) on the Keck I telescope \citep{VogtAllen1994}. CLS provides high-precision RV measurements for 719 FGKM stars and updated orbital solutions for known exoplanets, including GJ 436 b. Their earlier measurements (1996--2004) were from the California \& Carnegie Planet Search (CCPS; \citealt{CummingButler2008}), which was later split into the California Planet Search (CPS) and the Lick-Carnegie Exoplanet Survey (LCES). Since Keck/HIRES RV measurements reported by other studies \citep[e.g.][]{ButlerVogt2004,KnutsonFulton2014} fall within the time span covered by \cite{RosenthalFulton2021}, we adopt the CLS measurements, which represent the most recent version of standard CPS procedures \citep{HowardJohnson2010}. Due to a significant instrumental upgrade to the HIRES spectrograph in 2004, we treat pre-upgrade (HIRES1; 1996--2004) and post-upgrade (HIRES2; 2004--2019) measurements as two separate datasets with median RV uncertainties of 2.7~m~s$^{-1}$ and 1.6~m~s$^{-1}$, respectively.

We also use 171 RV measurements from the High Accuracy
Radial-velocity Planet Searcher \citep[HARPS;][]{MayorPepe2003} spectrograph ($\lambda$ = 378--691~nm, $R$ = 115,000), spanning 4.2 years from 2006 to 2010 \citep{LanotteGillon2014}.
The measurements from \cite{LanotteGillon2014} begin with the standard HARPS Data Reduction Software (DRS) values. Differential RVs were then derived with a $\chi^2$-fitting routine with a high signal-to-noise template from merged science spectra. These observations achieve a median RV uncertainty of 1.1~m~s$^{-1}$ for GJ 436 b.

We also include 113 observations from the Calar Alto high-Resolution search for M dwarfs with Exoearths with Near-infrared and optical \'{E}chelle Spectrographs (CARMENES) dual-channel \'{e}chelle spectrograph at Calar Alto Observatory \citep{QuirrenbachAmado2016}, obtained in 2016 and reported by \cite{TrifonovKurster2018}. CARMENES began operations in 2016 and is optimized for M-dwarf observations, with coverage in both the visible ($\lambda$ = 520--960~nm, $R$ = 94,600) and near-infrared ($\lambda$ = 960--1710~nm, $R$ = 80,400) wavelength ranges. \cite{TrifonovKurster2018} presented results from observations of seven confirmed M dwarf targets, including GJ 436 b, to test the capabilities of the newly built CARMENES instrument in probing close M dwarf hosts. The RV measurements were derived using the \texttt{CARACAL} \citep{caballerogaurdia2016} and \texttt{SERVAL} \citep{ZechmeisterReiners2018} pipelines, achieving a median uncertainty of 1.1~m~s$^{-1}$.

We model the combined RV dataset of GJ 436 b using \texttt{RadVel} \citep{FultonPetigura2018}, which implements a Bayesian framework for Keplerian orbit fitting. \texttt{RadVel} samples the model posteriors via maximum a posteriori optimization and Markov Chain Monte Carlo (MCMC), implemented with the Python package \texttt{emcee} \citep{Foreman-MackeyFarr2019}. We fit for the orbital period $P_{\mathrm{b}}$, time of inferior conjunction $T_{\mathrm{conj,b}}$, logarithm of the RV semi-amplitude $\log K_{\mathrm{b}}$, and jump parameters $\sqrt{e_{\mathrm{b}}}\cos\omega_{\mathrm{b}}$ and 
$\sqrt{e_{\mathrm{b}}}\sin\omega_{\mathrm{b}}$, where $e_{\mathrm{b}}$ is the eccentricity and $\omega_{\mathrm{b}}$ is the argument of periastron. We also fit for RV zero point $\gamma$ and jitter terms $\sigma$ for each instrument, as well as a global linear term $\dot{\gamma}$ and a quadratic RV curvature term $\ddot{\gamma}$. The priors and best-fit values and $1\sigma$ uncertainties for each of these parameters are listed in \autoref{tab: radvel table}.

\subsection{Joint RV-Astrometry Fit \label{jointRVAstro}}

We perform a joint fit using both RV and astrometry to account for all available observational constraints. In an attempt to directly constrain the properties of a putative perturber, we utilize the Python package \texttt{orvara}, which models stellar orbits and their companions through joint fits of RVs, relative astrometry, and absolute astrometry. \texttt{orvara} implements direct joint fitting of absolute astrometry through automated retrieval of data from the HGCA, using the Hundred Thousand Orbit Fitter \citep[\texttt{htof;}][]{BrandtMichalik2021} to compute synthetic Hipparcos and Gaia catalog positions and proper motions \citep{BrandtDupuy2021}. Our joint fit of RVs and absolute astrometry yields no meaningful constraints on the perturber's orbital parameters. This is consistent with the results of \cite{LanotteGillon2014}, who concluded that current data support a single-planet system, finding no significant peaks above the 3$\sigma$ confidence level in their periodogram analysis of RV residuals. Furthermore, Gaia DR3 astrometry is consistent with a single-star solution for GJ 436, with a Renormalized Unit Weight Error (RUWE) value of 1.3 \citep{GaiaCollaborationVallenari2023}. RUWE values above 1.4 are predominately binary stars \citep{LindegrenHernandez2018,StassunTorres2021}.

As our joint RV and astrometry model does not conclusively yield an outer companion, any remaining perturber would need to be substellar at a relatively wide orbital separation. To explore the parameter space where such a companion would remain undetected given the available data, we use \texttt{ethraid} \citep{VanZandtPetigura2024}, a publicly available Python package that samples companion masses and semi-major axes. \texttt{ethraid} is oriented toward systems with limited phase coverage, making it particularly suitable for the GJ 436 case.

\begin{figure}
    \centering    \includegraphics[width=1\linewidth]{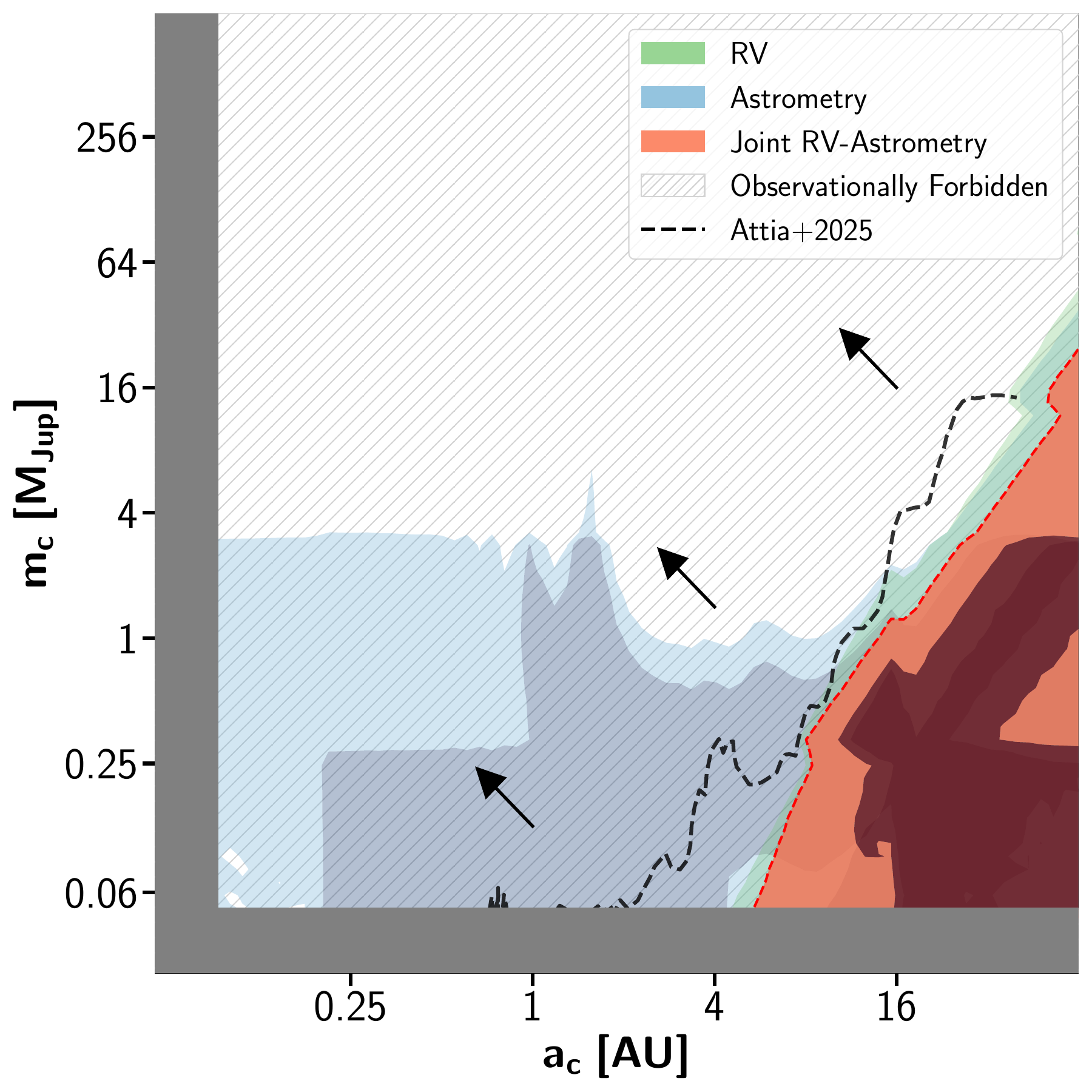}
    \caption{Joint posterior constraints in $m_\mathrm{c}$--$a_\mathrm{c}$ space. The dark and light regions indicate the $1\sigma$ and $2\sigma$ confidence intervals for  RV constraints (green), astrometric constraints (blue), and our final combined fit (red). The hashed region is observationally forbidden by our work. The parameter space above the black dashed lines is observationally forbidden by \Atwentyfive, see \S\ref{attia} for further discussion about this. For a more detailed comparison between this work and \Atwentyfive, see Section \ref{attia}. }
    \label{fig:param_plot}
\end{figure}

We examine the parameter space for an outer perturber with semi-major axes $a_\mathrm{c}$ ranging from 0.09--64~AU and masses $m_\mathrm{c}$ ranging from 0.05--1000~$M_\text{Jup}$. The lower mass bound of 0.05~$M_\text{Jup}$ was chosen to avoid a numerical instability in the \texttt{ethraid} code that occurs below this threshold, while the upper bound was set as the maximum mass available in \texttt{ethraid}. We employ 50 million orbital models distributed over a $70\times70$ grid to assess companion probabilities over the entire parameter space, using the measured $\dot{\gamma}$ and $\ddot{\gamma}$ as reported in \autoref{tab: radvel table}. We adopt the \texttt{piecewise} prior in \texttt{ethraid}, which applies the \cite{Kipping2013} distribution for planetary masses ($m_\mathrm{c} < 13~M_{\text{Jup}}$), the Beta distribution from \cite{BowlerBlunt2020} for brown dwarfs ($13~M_{\text{Jup}} < m_\mathrm{c} \le 80~M_{\text{Jup}}$), and a uniform distribution between $0.1M_{\odot}$ and $0.8$~$M_{\odot}$ for stars ($m_\mathrm{c} > 80~M_{\text{Jup}}$) from \cite{RaghavanMcAlister2010}. All other priors were set to the \texttt{ethraid} defaults: uniform distributions for $\log(a_\mathrm{c}/1\text{AU}$), $\log(m_\mathrm{c}/M_{\text{Jup}}$), cosine of the orbital inclination ($\cos{i}$), argument of periastron ($\omega$), and mean anomaly at the reference epoch ($M$). \autoref{fig:param_plot} presents the resulting joint posterior constraints in $m_\mathrm{c}$--$a_\mathrm{c}$ parameter space, where the RV constraint, astrometric constraint, and joint constraint are shown as green, blue, and red contours, respectively. At the $2\sigma$ confidence level, the joint fit limits potential companions to $a_{\mathrm{c}}<5.4$~AU for $m_{\mathrm{c}}>0.05~M_{\mathrm{Jup}}$ and $a_{\mathrm{c}}<64$~AU for $m_{\mathrm{c}}>24~ M_{\mathrm{Jup}}$ within the examined parameter space. 

\begin{figure*}
    \centering
    \includegraphics[width=1\linewidth]
    {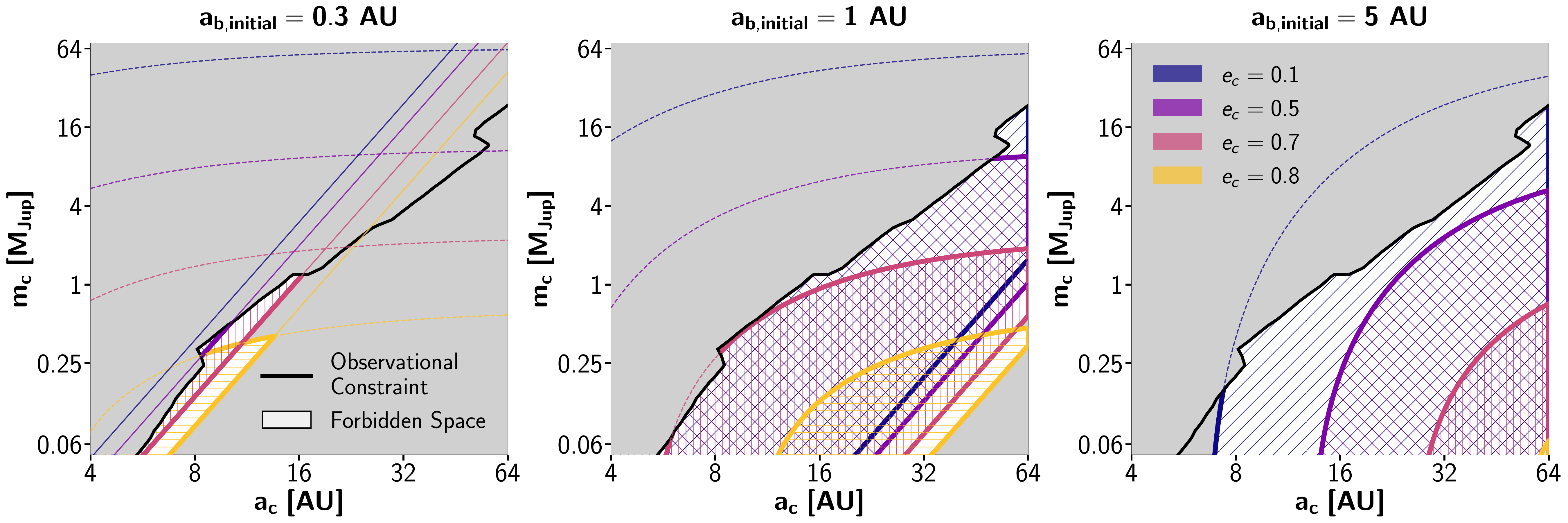}
    \caption{Analytic constraints in $m_\mathrm{c}$--$a_\mathrm{c}$ space for three initial semi-major axes $a_{\mathrm b} = 0.3, 1, 5$ AU. Light gray shading with no overlaid, colored hatch pattern indicates the observationally ($2\sigma$) and dynamically forbidden region. The solid black line delineates the observationally forbidden region. Each color corresponds to a different outer companion eccentricity $e_{\mathrm c} = 0.1$ (dark purple), $e_{\mathrm c} = 0.5$ (violet), $e_{\mathrm c} = 0.7$ (pink), and $e_{\mathrm c} = 0.8$ (orange). The colored solid lines delineate where $\frac{t_{\mathrm{GR}}}{t_{\mathrm{quad}}} = 1$, above which ZLK precession dominates over GR precession. The dashed curves indicate Hill stability boundaries with a separation criterion $\Delta = 5$, below which the system remains dynamically stable. Colored regions represent the parameter space where the outer perturber satisfies all three criteria: (1) dynamically stable (below the dashed curve), (2) ZLK precession dominates over GR analytically (above the solid line), and (3) observationally allowed.}
    \label{fig:analytic_plot}
\end{figure*}

\section{Theoretical Background \label{theoretical background}}

As discussed in \S\ref{introduction}, the most dynamically puzzling aspect of this system is the residual orbital eccentricity of GJ 436 b. Using a standard tidal quality factor for a Neptune-like planet, $Q = 10^5$ (e.g., \citealt{millholland_2019, LuAn2025}), the tidal circularization time of GJ 436 b is estimated to first order as \citep{goldreich1966}:
\begin{equation}
\label{eq:tcirc}
t_{\text{circ}} = \frac{4}{63} \frac{a_\mathrm{b}^{13/2}}{\sqrt{GM_*^3}} Q m_\mathrm{b} R_\mathrm{b}^{-5} \sim 2 \times 10^8 \text{ years},
\end{equation}
where $G$ is the gravitational constant, $M_*$ is the stellar mass, and $m_\mathrm{b}$, $R_\mathrm{b}$, and $a_\mathrm{b}$ are the mass, radius, and semi-major axis of GJ 436 b, respectively. The estimated age of the system exceeds the expected circularization timescale: \cite{VeyetteMuirhead2018} derived an age of $8.9^{+2.3}_{-2.1}$ Gyr from kinematics and $\alpha$-element enrichment, while \cite{BourrierLovis2018} estimated an age of 4--8 Gyr based on gyrochronology. Earlier work by \cite{ButlerVogt2004} also concluded that the system is older than 3 Gyr from kinematic and chromospheric diagnostics. Given that the age estimates for this system are unanimously projected to be several Gyr, GJ 436 b should have already undergone complete tidal circularization if it had formed in situ.\footnote{Note that these first-order analytic expressions tend to overestimate the circularization timescale \citep[e.g.][]{wisdom_2008}, so our conclusion that GJ 436 b should be completely circularized is even stronger than the numbers may suggest.
} Its residual eccentricity of $e\sim0.16$ is thus a puzzle that merits investigation. 

\begin{figure*}
    \centering
    \includegraphics[width=\textwidth]{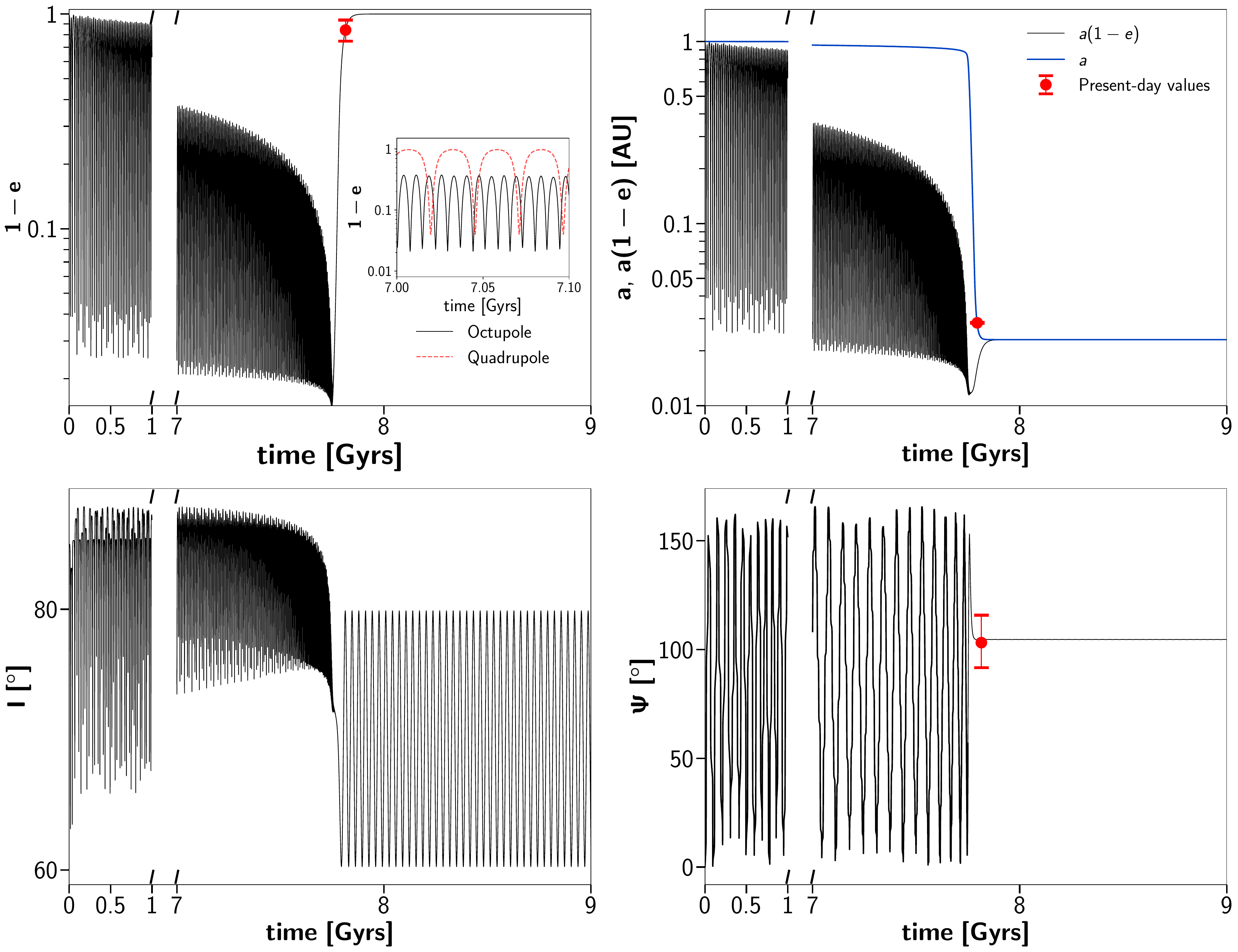}
    \caption{Example dynamical history of GJ 436 b via ZLK migration with an outer companion of mass $m_\mathrm{c} = 0.24~M_\text{Jup}$ on an orbit with $a_\mathrm{c} = 28$ AU, $e_\mathrm{c} = 0.62$, and $I_{\mathrm{mut,i}} = 85^\circ$. GJ 436 b was initialized with $a_{\mathrm{b,i}}=1$~AU. The panels show the evolution of $1-e$ (top left), mutual inclination $I$ (bottom left), semi-major axis $a$ and periapse distance $a(1-e)$ (top right), and spin-orbit angle $\Psi_{\mathrm{Ab}}$ (bottom right). The present-day state captures GJ 436 b during its ongoing tidal evolution before reaching the final semi-major axis plateau, as the planet has not yet achieved complete tidal circularization given its observed residual eccentricity. The red points indicate the present-day values for each parameter with measurement uncertainties. The time axis is truncated between 1--7~Gyr, indicated by break marks. The inset in the top left panel compares the octupole-level (black solid line) versus quadrupole-level (red dashed line) approximation. }
    \label{fig:success_plot}
\end{figure*}

\begin{figure*}
    \centering
    \includegraphics[width=1\linewidth]
    {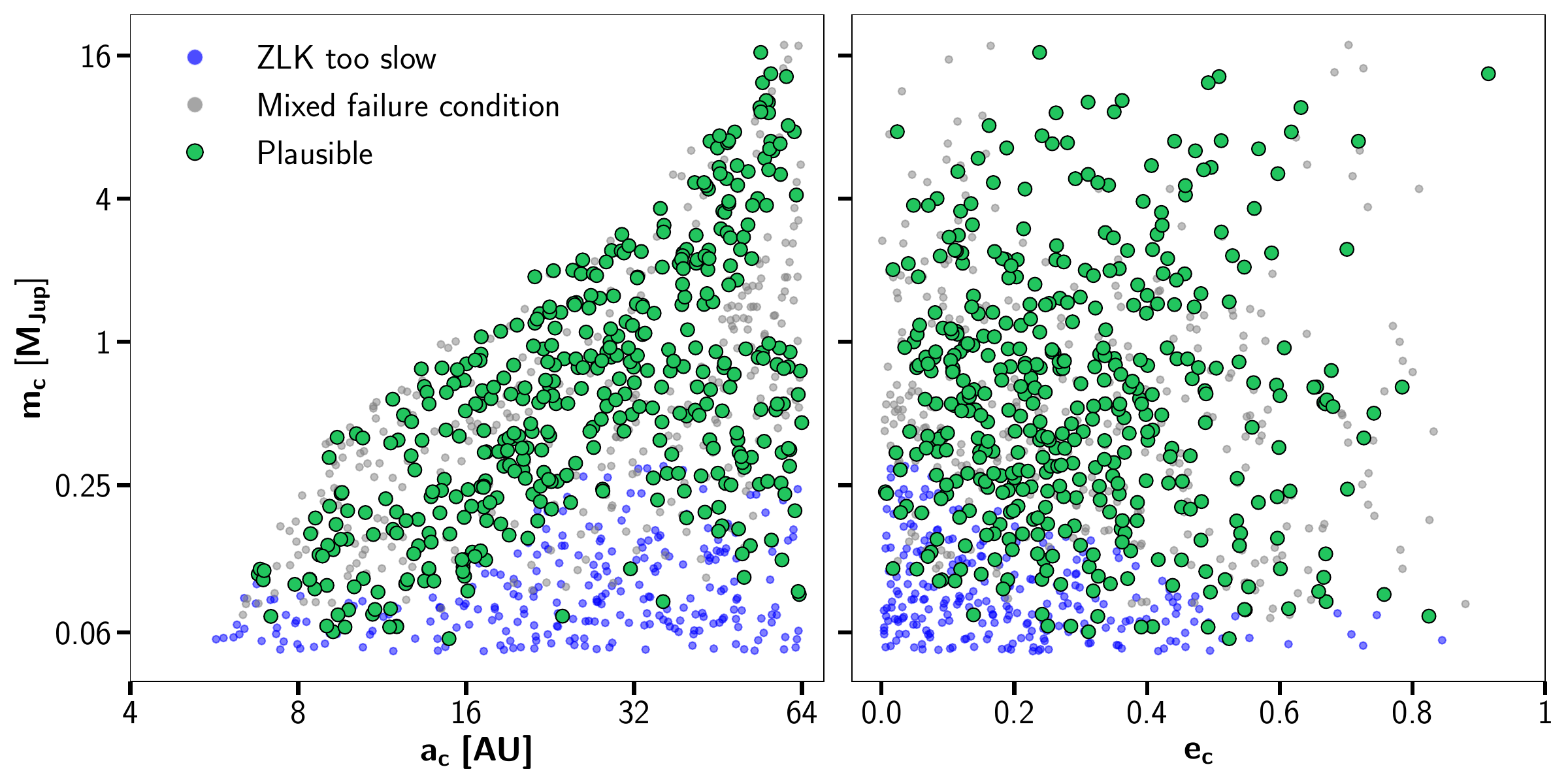}
    \caption{ZLK simulation results in $a_\mathrm{c}-m_\mathrm{c}$ (left) and $m_\mathrm{c}-e_\mathrm{c}$ space (right). Each point represents one companion configuration tested across 18 different initial conditions. Blue points indicate companion configurations where ZLK migration is too slow ($t_\mathrm{HN} > $ 10 Gyr) for all 18 initial conditions. Gray points show mixed failure conditions where ZLK migration either is too slow ($t_\mathrm{HN} > $ 10 Gyr), is too fast ($t_\mathrm{HN} < $ 1 Gyr), or causes GJ 436 b to reach the Roche limit ($a_b<a_R$). Green points represent plausible configurations for which at least one of the 18 suites reproduces the present-day orbit. For each plausible companion configuration, only one to three of the 18 initial condition suites plausibly reproduce the present-day orbit.}
    \label{fig:sims_plot}
\end{figure*}

\subsection{Dynamical Formation Scenarios Without a Present-Day Perturber \label{alternative_mechanisms}}

Given our inability to detect a companion with decades of available radial velocity and astrometric data, we must address what would seem to be the simplest explanation: perhaps there simply is no perturber. We argue that scenarios that do not require a still-present perturber have great difficulty in simultaneously explaining GJ 436 b's eccentricity and spin-orbit misalignment. While these scenarios cannot be completely ruled out, they require a confluence of low-likelihood events.

One possible avenue to form GJ 436 b with no additional perturber is through planet-disk eccentricity excitation within a polar-aligned protoplanetary disk. Previous studies have shown that planet-disk interactions are capable of generating modest primordial eccentricities on the order of GJ 436 b's current value \citep[e.g.][]{duffell2015eccentric}. However, such models struggle to retain the currently observed elevated eccentricity of GJ 436 b over Gyr timescales. Furthermore, while a polar primordial disk misalignment is not impossible, it is challenging to invoke in the GJ 436 system: the protoplanetary disk may be readily tilted by a stellar companion, for example \citep[e.g.][]{spalding_2014, zanazzi_misalignment,gerbig_2024, SuLai2025}, but there is no evidence supporting the existence of a stellar companion. We therefore conclude that disk interactions alone are unlikely to account for both the orbital eccentricity and spin-orbit misalignment of GJ 436 b.

Alternatively, planet-planet scattering can endow planets with significant eccentricity and spin-orbit misalignment while completely ejecting the companion responsible from the system \citep[e.g.][]{chatterjee_2008, nagasawa2008scattering, juric_tremaine_2008, carrera2019planet}. However, planet-planet scattering typically occurs shortly after the protoplanetary disk dissipates, when systems often undergo instabilities \citep{MoorheadAdams2005, MatsumuraTakeda2008, PichierriBitsch2023}. For an M dwarf such as GJ 436, the median disk lifetime is estimated to be 5--10 Myr \citep{PfalznerDehghani2022,PfalznerDincer2024}. Thus, considering the age of GJ 436, any significant eccentricity endowed by scattering would have to withstand several Gyr of tidal damping, which is unlikely for a standard Neptune-like $Q$ value.\footnote{If GJ 436 b is an order of magnitude less dissipative than one would expect from a Neptune-like planet, as posited by \citet{mardling_2008}, the planet's circularization timescale would approach the age of the system, and thus an orbital eccentricity may be feasibly maintained. While unlikely, this scenario cannot be definitively excluded given the upper bounds of $Q'$ estimates from \cite{MorleyKnutson2017}.} A late scattering event, in which the system remains stable over several Gyr after the dissipation of the disk before going unstable, is feasible but would require an external trigger such as a stellar flyby \citep{malmberg2011flyby} or two or more extremely widely-spaced companion planets with long stability timescales \citep[e.g.][]{Chambers_1996, obertas_2017, denham_2019,lammers2024instability}.

\subsection{Dynamical Formation Scenarios With a Present-Day Perturber}
The presence of an undetected perturber in the GJ 436 system opens up a range of alternative possible formation pathways. Numerous authors, including \cite{RibasFont-Ribera2008}, \cite{batygin2009gj436}, and \cite{petrovich_2020}, have proposed secular dynamical evolutionary pathways involving planetary perturbers that would remain bound to the system. However, the perturbers necessary for these solutions fall within parameter space forbidden by our observational constraints, as delineated in \autoref{fig:param_plot}.

We hence posit that ZLK-driven tidal migration provides the most plausible explanation for GJ 436 b's orbit. Numerous studies have explored ZLK migration in the GJ 436 system.  \citet{beust_2012} first applied secular simulations to demonstrate that starting GJ 436 b 5--10 times more distant than its present-day orbit could enable the preservation of GJ 436 b's orbital eccentricity over 10 Gyr. They explored the parameter space of companion masses and orbital periods, finding that successful ZLK migration requires highly inclined configurations within the observational and analytical limits. \citet{BourrierLovis2018} updated the viable parameter space of companion masses and orbital periods to a narrow region, assuming $a_{\mathrm{b,i}} = 0.35$ AU and $I_{\mathrm{mut,i}} = 85^{\circ}$. \citet{attia2021JADE} modeled ZLK migration in the GJ 436 system accounting for coupled atmospheric evolution with their \texttt{JADE} code. Recently, \Atwentyfive refined the \texttt{JADE} code and explored the parameter space of companion masses and orbital periods using a grid method, ultimately concluding that GJ 436 b should have likely started with $a_{\mathrm{b,i}}\sim0.3$ AU. 

Our study improves on these works in two key ways. First, our observational constraints on potential perturber parameter space are more extensive. Second, we include methodological improvements in our dynamical analysis, elaborated upon in \S \ref{attia}.

\subsection{Theoretical Background on ZLK migration} \label{ZLK migration subsec}
ZLK migration is an appealing explanation for the present-day orbit of GJ 436 b, given its ability to reproduce both the planet's polar orbit and residual eccentricity without assuming non-standard tidal $Q$ values for Neptune-like planets. In three-body hierarchical systems with a significant mutual inclination between the orbits ($I_\mathrm{mut}$), the outer perturber can periodically excite the eccentricity and inclination of the inner body's orbit \citep{von_zeipel_1910, lidov1962evolution, kozai1962secular, naoz2016eccentric}. During each cycle of maximum eccentricity, the inner planet undergoes strong tidal forces at periastron passages such that orbital energy is dissipated; thus the semi-major axis of the planet shrinks. This process is known as ZLK migration, and it has emerged as a promising formation mechanism for hot Jupiters and other close-in planets \citep[e.g.,][]{wu_2003, fabrycky2007shrinking, naoz_2011, naoz_2012, Dong2014, Anderson2016, Stephan_2018, Dawson2018, rice2022origins..17R,LuAn2025,liu2025formation, liu2026wasp94}. 

To the quadrupole order, the ZLK precession rate can be expressed as \citep{naoz2013, naoz2016eccentric, LuAn2025}

\begin{equation}
t_{\text{quad}} \sim \frac{16 a_\mathrm{c}^3 (1-e_\mathrm{c}^2)^{3/2} \sqrt{M_* + m_\mathrm{b}}}{15 a_\mathrm{b}^{3/2} m_\mathrm{c} \sqrt{G}},
\end{equation}

\noindent where $a_\mathrm{c}$, $m_\mathrm{c}$, and $e_\mathrm{c}$ are the semi-major axis, planetary mass, and eccentricity of the outer companion's orbit, respectively. 

As the orbit shrinks, ZLK oscillations are suppressed due to short-range forces: General Relativity (GR), tides, and rotational distortions. In particular, GR effects can induce periapsis precession in the opposite direction of the ZLK precession, where the ratio of the GR precession timescale to the quadrupole ZLK timescale is given by  \citep{naoz2013, naoz2016eccentric, LuAn2025}

\begin{equation}
\label{eq: GR-ZLK-ratio}
\frac{t_{\mathrm{GR}}}{t_{\mathrm{quad}}} \sim \frac{a_\mathrm{b}^4}{3 a_{\mathrm{c}}^3} \frac{\left(1-e_\mathrm{b}^2\right) m_\mathrm{c} c^2}{\left(1-{e_\mathrm{c}}^2\right)^{3 / 2}G\left(M_{\star}+m_\mathrm{b}\right)^{2}}.
\end{equation}

Using the present-day parameters of GJ 436, we find $t_{\mathrm{GR}}/{t_{\mathrm{quad}}}<1$, indicating that GR precession dominates over ZLK. Thus, the eccentricity of GJ 436 b cannot be attributed to presently-ongoing ZLK oscillations. However, ZLK migration may have occurred in the past history of the system, despite having been since suppressed.

\subsection{Analytic constraints}

In this section, we leverage analytic constraints to delineate the regions of parameter space where ZLK migration is a feasible mechanism for generating GJ 436 b's current orbit. In \autoref{fig:analytic_plot}, we re-plot the observationally allowed parameter space from \autoref{fig:param_plot} together with our further constraints from analytic arguments that compare GR precession and ZLK precession timescales and impose dynamical stability conditions. These conditions are described in greater detail below. 

These arguments necessitate some assumption regarding the initial orbit of GJ 436 b. We evaluate these constraints for three different $a_{\rm b,i} = 0.3, 1, 5$ AU, displayed in the left, center, and right panels of \autoref{fig:analytic_plot}, respectively. The shaded regions with different colors represent the dynamically allowed parameter space for different companion eccentricities $e_{\mathrm{c}} = 0.1$, 0.5, 0.7, and 0.8.

\begin{enumerate}[label=(\alph*)]

    \item \textit{GR precession}: When the GR precession timescale is shorter than the ZLK precession timescale, then ZLK oscillations are suppressed. Setting the ratio of these timescales to 1 (see \autoref{eq: GR-ZLK-ratio}) defines the boundary where GR precession becomes dominant. In \autoref{fig:analytic_plot}, the colored solid lines in each panel delineate ${t_{\mathrm{GR}}}/{t_{\mathrm{quad}}} = 1$, above which ZLK precession dominates. Beyond these lines, the perturber is either too low-mass or too distant to generate enough precession to overcome GR effects, and hence ZLK should not begin at all.

    \item \textit{Dynamical Stability}: For a two-planet system with low-eccentricity orbits and planetary masses much lower than the host star's mass, the Hill stability criterion provides an analytic condition for the minimum orbital spacing between the planets that is required for long-term dynamical stability. This critical separation is conventionally expressed in units of mutual Hill radius, defined as \citep{Gladman93}
    
    \begin{equation}
        R_H = \frac{a_{\mathrm{b}} + a_{\mathrm{c}}}{2} \left(\frac{m_{\mathrm{b}} + m_{\mathrm{c}}}{3 M_*}\right)^{1/3}. 
    \end{equation}
    To account for orbital eccentricity, we define the minimum physical separation $\Delta$ using 
    
    \begin{equation}
        \Delta = \frac{a_{\mathrm{c}}(1-e_{\mathrm{c}}) - a_{\mathrm{b}}(1+e_{\mathrm{b}})}{R_H}.
    \end{equation}

    We arbitrarily choose $\Delta=5$ as the criterion for dynamical stability, given that a wide range of values, from 3 to 10, is used in the literature \citep{Gladman93,Chambers_1996}. In \autoref{fig:analytic_plot}, the colored dashed lines delineate the dynamical stability boundary for four different outer companion eccentricities, below which the system becomes dynamically stable. Above the boundary, the perturber is either too massive or too close to be dynamically stable. If a larger $\Delta$ is selected, the dashed lines would be pushed rightward since a larger semi-major axis would be required for stability.

    \item \textit{Flip and descent timescales}: The analytic flip timescale is the time required for the inner planet to first flip from $I_{\mathrm{mut}} < 90^{\circ}$ to $I_{\mathrm{mut}} > 90^{\circ}$ assuming an eccentric inner orbit ($e_1 > 0.6$) on an initially coplanar configuration with a perturber on an eccentric outer orbit (see Equation 9 of \citealt{li_eccentricity_2014} and Equation 36 of \citealt{naoz2016eccentric}). This timescale indicates how quickly the inner planet can reach extremely high eccentricities through octupole-order effects. In our explored parameter space of $a_{\mathrm{c}}$--$m_{\mathrm{c}}$, the flip timescales are significantly shorter than the expected age of the system, providing no meaningful constraints on the companion properties. Similarly, we calculate the secular descent time (see Equation 34 of \citealt{weldon2024descent}), which is another indicator of the inner planet reaching high eccentricities. However, the descent timescale provides weaker constraints than those imposed by GR precession. 
\end{enumerate}

The viable parameter space shown in \autoref{fig:analytic_plot} consists of the overlapping regions that simultaneously satisfy all three constraints: ZLK precession dominates over GR precession (above solid lines), the systems are dynamically stable (below the dashed lines), and the companion is consistent with the observational constraints in \autoref{fig:param_plot}. For $a_{\mathrm{b,i}} = 0.3$ AU (left panel), the analytic constraints are extremely restrictive and rule out the vast majority of parameter space, including most of the parameter space considered in \Atwentyfive. Hence, the majority of successful simulations in \Atwentyfive should not have begun ZLK oscillations. Indeed, only two of the $1000$ companion configurations we ran with $a_{\mathrm{b,i}} = 0.3$ AU plausibly reproduce the present-day orbit, compared to 329 and 95 for $a_{\textrm{b,i}}=1$ AU and $a_{\textrm{b,i}}=5$ AU, respectively (see \S\ref{simresults} for further details).

\section{Dynamical Simulations \label{dynamical_sims}}

In this section, we run a series of secular dynamics simulations to place further constraints on the parameter space of the potential GJ 436 c companion and demonstrate that ZLK migration can reproduce the present-day orbit of the planet.

\subsection{Dynamical Simulation Setup} \label{simulation_setup}

ZLK oscillations require a large initial mutual inclination between planetary orbits, and our constraints on GJ 436 c cannot rule out an eccentric orbit. Neither is naively expected from in situ formation. Extensive previous work has shown that strong planet-planet scattering can place the outer planet on an inclined orbit, setting initial conditions favorable to triggering ZLK oscillations \citep[e.g.][]{beauge2012multiple,LuAn2025,weldon2025cold}, while typically producing elevated eccentricities as a side effect. Therefore, we do not explicitly simulate the scattering process, but instead place GJ 436 c on an initially inclined orbit and allow for a range of eccentricities. 

Building on the observational constraints derived in \S\ref{jointRVAstro}, we simulate the dynamical evolution of GJ 436 b using secular simulations to further narrow the parameter space of the perturbing companion. We use \texttt{KozaiPy},\footnote{https://github.com/djmunoz/kozaipy} a publicly available Python code that integrates secular equations of motion at the octupole level of approximation from \cite{eggleton_2001} and \cite{fabrycky2007shrinking} for three-body hierarchical systems. Short-range forces \citep{naoz2013,li_eccentricity_2014} are included, taking into account precession due to general relativity, rotational flattening, and tidal distortion consistent with the constant time lag model of equilibrium tides \citep[e.g.][]{alexander1973weak,hut1981tidal,eggleton1998equilibrium, mardling2002calculating}.

We draw 1,000 random samples of self-consistent $a_\mathrm{c}$, $m_\mathrm{c}$, and $e_\mathrm{c}$ combinations from the joint fit posterior shown in \autoref{fig:param_plot} to set up the orbit of GJ 436 c for the simulations. For each companion configuration, we run 18 simulation suites covering all combinations of inner planet initial semi-major axis $a_{\textrm{b,i}} \in \{0.3,~1,~5\}$ AU and initial mutual inclination $I_{\mathrm{mut,i}} \in \{45,~60,~75,~80,~85,~89\}^\circ$. The initial conditions for GJ 436 b are necessarily arbitrary as the planet's formation location is unconstrained. All other variables are fixed as follows. We set the initial eccentricity of the planet to $e_{\text{b,i}} =0.01$ and adopt arbitrary values for the argument of periapsis for the inner and outer planets of $\omega_b = 0^\circ$ and $\omega_c = 45^\circ$. We set the apsidal constant for the star and planet to $k_{s}=0.014$ and $k_{b}=0.25$, respectively, and a stellar viscous timescale of $t_{v,s}= 55$ years following \cite{fabrycky2007shrinking}. The planetary viscous timescale is set to $t_{v,b}= 0.75$ years, derived using the observed reduced tidal quality factor $Q' = 2 \times 10^5$ \citep{MorleyKnutson2017} and the present-day semi-major axis via the following relation \citep{fabrycky2007shrinking}:

\begin{equation}
Q = \frac{4}{3} \frac{k_{b}}{(1 + 2k_{b})^2} \frac{G m_\mathrm{b}}{R_\mathrm{b}^3} \left( \frac{a_\mathrm{b}^3}{GM_*} \right)^{1/2} t_{v,b}.
\end{equation}

\subsection{Dynamical Simulation Results} \label{simresults}
\autoref{fig:success_plot} presents a fiducial ZLK migration track for GJ 436 b. This simulation includes an outer companion of mass $m_\mathrm{c} = 0.24~M_\text{Jup}$ initialized on an orbit with $a_\mathrm{c} = 28$ AU, $e_\mathrm{c} = 0.62$, and $I_{\mathrm{mut,i}} = 85^\circ$. The simulation shows the characteristic features of ZLK evolution: periodic eccentricity and inclination oscillations. The planet's semi-major axis decreases gradually due to tidal dissipation until ZLK effects are quenched and tidal circularization begins at approximately 7 Gyr. The spin-orbit angle exhibits large variations ranging from $0^\circ$ to $170^\circ$ before stabilizing at $105^\circ$, within $1\sigma$ of the spin-orbit angle measured by \cite{BourrierZapateroOsorio2022}. The planet reaches a final semi-major axis of $\sim 0.023$ AU upon full tidal circularization. While this is smaller than the final semi-major axis expected from present-day observations, the discrepancy can be reconciled by incorporating the expected influence of radius inflation (see \autoref{appendix}). This demonstrates that high-eccentricity migration via ZLK oscillations can reproduce the present-day orbit of GJ 436 b.

We present the population-level results of the 18,000 simulations for 1,000 companion configurations drawn from the joint posterior in \autoref{fig:sims_plot}. We define a system to have plausibly reproduced the present-day orbit of GJ 436 b if $a_\mathrm{b} < 0.05$~AU. We do not attempt to exactly match the observed semi-major axis of GJ 436 b, as the chaotic nature of the system makes it extremely challenging to reproduce the precise observed value. A qualitative match is sufficient to demonstrate the viability of the ZLK mechanism. 

\begin{figure}
    \centering
    \includegraphics[width=1\linewidth]{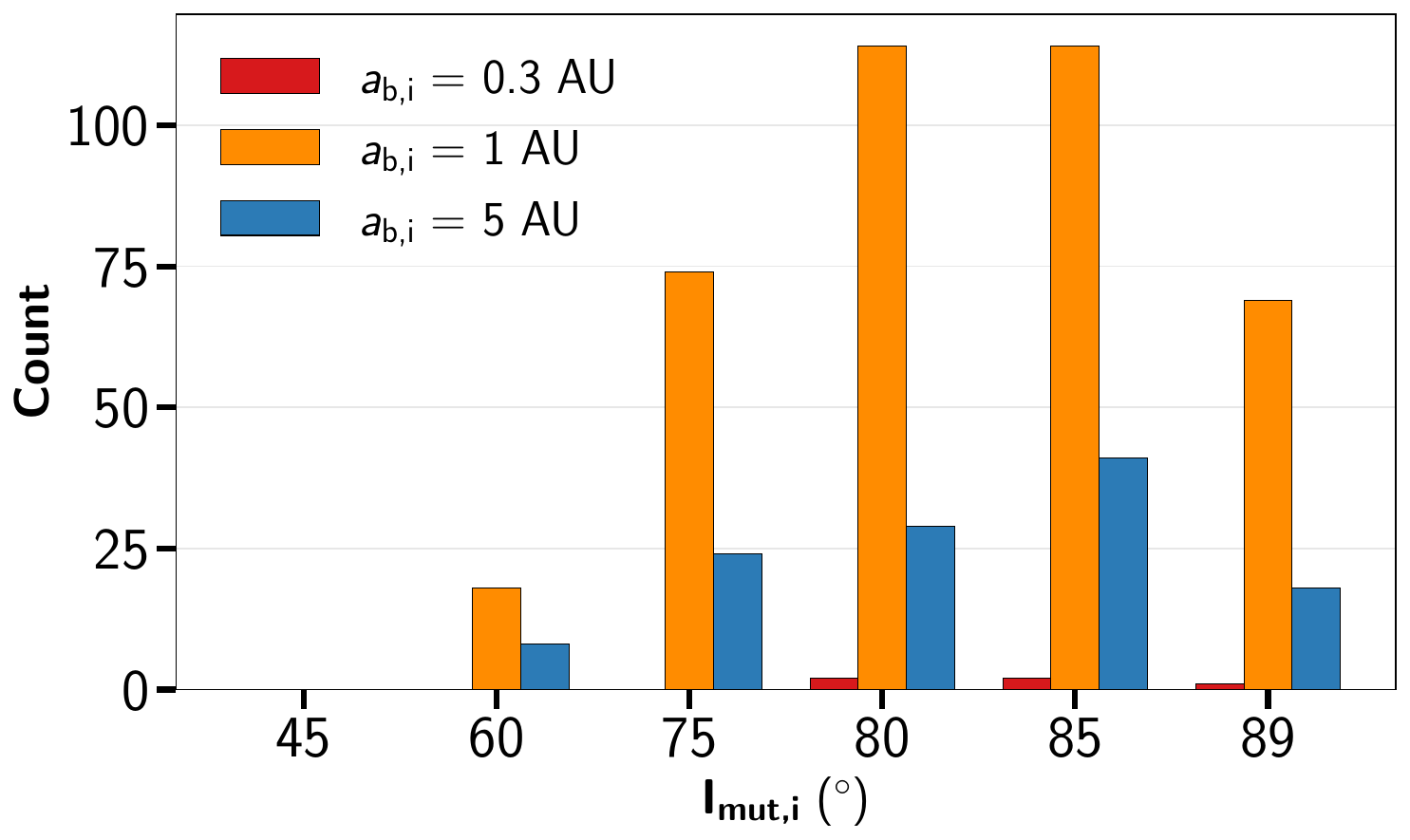}
    \caption{Number of plausible simulations versus initial mutual inclination $I_{\mathrm{mut,i}}$ for three initial semi-major axes $a_{\mathrm{b,i}} = 0.3$ AU (red), 1 AU (orange), 5 AU (blue). None of the simulations initialized with $I_{\mathrm{mut,i}}=45^\circ$ and very few initialized with $a_{\mathrm{b,i}} = 0.3$ AU plausibly reproduce the present-day orbit. }
    \label{fig:inc_plot}
\end{figure}

\begin{figure}
    \centering
    \includegraphics[width=1\linewidth]{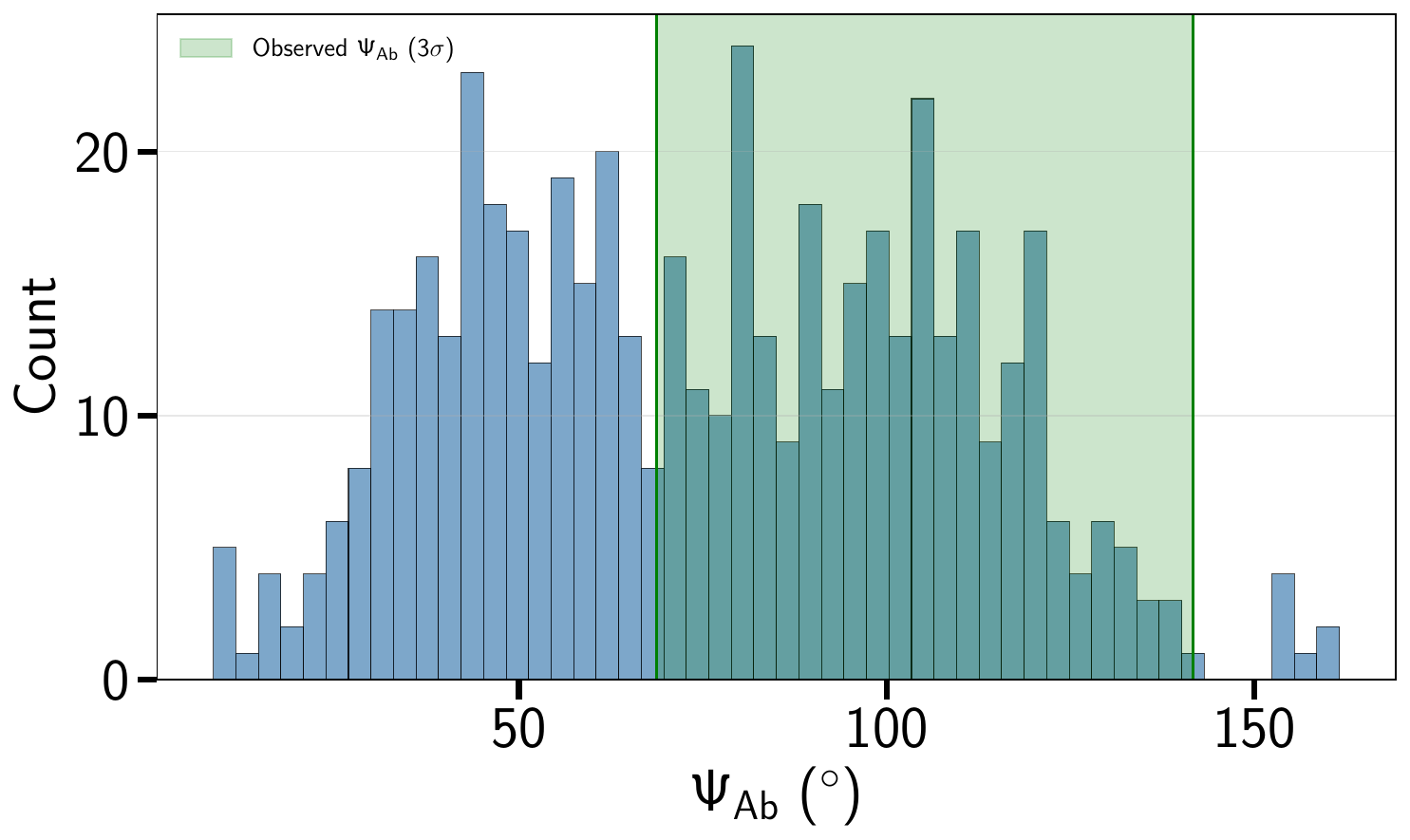}
    \caption{Distribution of spin-orbit angles $\Psi_{\mathrm{Ab}}$ for simulations that plausibly reproduce the present-day orbit (1 Gyr $< t_\mathrm{HN} < $ 10 Gyr) among the 18,000 simulations. The green shaded region indicates the $3\sigma$ confidence interval for the observed spin-orbit angle $\Psi_{\mathrm{Ab}} = {103.2}_{-11.5}^{+12.6}{}\degree$ \citep{BourrierZapateroOsorio2022}.}
    \label{fig:obliquity_plot}
\end{figure}

We categorize companion configurations according to the timescale taken to reach the present-day orbit, which we refer to as $t_{\rm HN}$. This timescale is defined as the sum of the time taken to quench ZLK oscillations and the subsequent tidal circularization timescale, thus representing the total time taken to evolve GJ 436 b from a distant orbit to its present-day close-in near-circular configuration. Each companion configuration is placed into one of the following three categories based on the collective behavior of its 18 simulations with varying initial conditions:

\begin{enumerate}[label=(\alph*)]

\item \textit{ZLK too slow}: All 18 suites fail to reproduce the present-day orbit because $t_\mathrm{HN} > $ 10 Gyr. In these simulations,  GJ 436 b is unable to migrate to a close-in orbit within the age of the system.

\item \textit{Mixed failure conditions}: All 18 suites fail to reproduce the present-day orbit because either (1) $t_\mathrm{HN} > $ 10 Gyr; (2) $t_\mathrm{HN} < $ 1 Gyr, where GJ 436 b reaches its present-day orbit early enough in its lifetime that any residual eccentricity would have damped out by the present day; or (3) GJ 436 b reaches the Roche limit \citep{MatsumuraPeale2010}, defined as

\begin{equation}
    a_{\mathrm{R}}= \frac{3}{2}R_{\mathrm b} \left[\frac{m_{\mathrm b}}{3(M_* + m_{\mathrm b})}\right]^{-1/3},
\end{equation}
at which we assume the planet is tidally disrupted.

\item \textit{Plausible}: At least one suite succeeds in reproducing the present-day orbit among the 18 initial condition tests, where 1 Gyr $< t_\mathrm{HN} < $ 10 Gyr is considered plausible. In these simulations, GJ 436 b is migrated to a close-in orbit sufficiently late such that it is possible to maintain some present-day residual eccentricity.
\end{enumerate}
These three categories contain 25.8\%, 32.5\%, and 41.7\% of companion configurations, respectively. 0.65\% of the 18,000 simulations reach the Roche limit. None of the simulations initialized with $I_{\mathrm{mut,i}}=45^\circ$ plausibly reproduce the present-day orbit (see \autoref{fig:inc_plot}). In \autoref{fig:obliquity_plot}, we plot the distribution of the median spin-orbit angle $\Psi_{\mathrm{Ab}}$ after the planet has reached $t>t_{\mathrm {HN}}$ for simulations that fall into the plausible category. 53.7\% of $\Psi_{\mathrm{Ab}}$ fall within the present-day $3\sigma$ observational constraint. 

Only 0.2\% of the 1,000 companion configurations initialized with $a_\mathrm{b} = 0.3$ AU reproduce the present-day orbit, while the rest of the plausible cases are initialized with $a_\mathrm{b,i} =1$~AU or $5$~AU. We conclude that it is unlikely that GJ 436 b formed at $\sim 0.3$ AU, as further supported by our analytic constraints shown in the left panel of \autoref{fig:analytic_plot}. This is consistent with young giant planet surveys, which find that giant planet occurrence rates inside 2.5~AU increase or remain constant over time \citep{GrandjeanLagrange2023, TranBowler2025} --- implying that close-in giant planets may typically migrate inward at late times. 

Plausible ZLK migration cases yield companion masses of $0.55_{-0.39}^{+1.67}~M_{\text{Jup}}$ (16th--84th percentile), suggesting a substellar companion. The associated semi-major axis range of $27.4_{-13.7}^{+21.8}$ AU indicates that the companion likely either formed beyond the ice line or migrated there post-formation. Companion configurations that exhibit slow ZLK migration regardless of $a_{\textrm {b,i}}$ (``ZLK too slow") have  $m_{\mathrm{c}}\lesssim 0.30~M_{\text{Jup}}$. While the $a_{\mathrm{b,final}}$ across our plausible simulations do not exactly match the value expected from present day observations, we emphasize our systems are qualitatively very similar and consider the lack of exact match to the present-day semi-major axis a fine-tuning issue. We offer dynamical evolution coupled with tidally-driven radius inflation as one possible solution to this discrepancy, and elaborate in detail in \autoref{appendix}.

\begin{figure}
    \centering
    \includegraphics[width=1\linewidth]{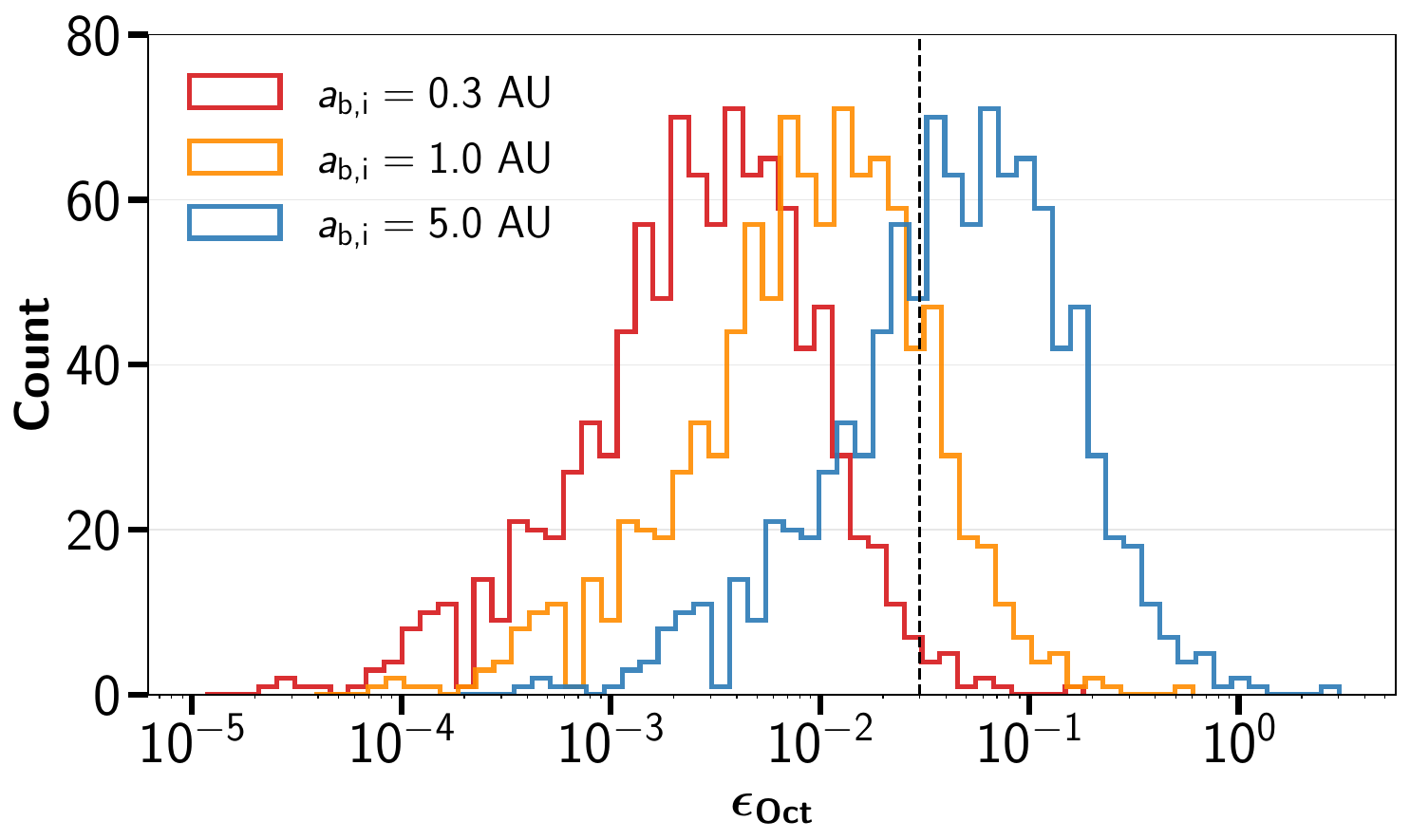}
    \caption{Distribution of $\epsilon_{\mathrm{Oct}}$ across 1,000 different companion configurations for three initial semi-major axes $a_{\mathrm{b,i}}=0.3$ AU (red), 1 AU (orange), 5 AU (blue). The black dashed line marks the $\epsilon_{\mathrm{Oct}}=0.03$, above which octupole-order effects become very significant for $I_{\mathrm{mut,i}}>60^\circ$.  
    A substantial fraction of our configurations exceed this threshold. }
    \label{fig:epsilon_oct}
\end{figure}

\section{Comparison to Attia et al. (2025)}
\label{attia}

We compare our methodology and results to those of \Atwentyfive in this section. Both studies converge on the key conclusion that ZLK migration driven by a substellar-mass companion can plausibly explain the present-day orbit of GJ436 b. However, while \Atwentyfive reports that GJ 436 b likely formed at $\sim$0.3~AU, only 0.2\% of our companion configurations initialized with $a_\mathrm{b,i} =0.3$~AU plausibly reproduce the present-day orbit of GJ 436 b. We note that \Atwentyfive place greater emphasis on reproducing the present-day semi-major axis, which motivates their adopted value of $a_\text{b, i} = 0.3$~AU. While we place less emphasis on exactly matching the present-day $a_{\text{b,final}}$, we argue that attempting to exactly match the extremely well-constrained present-day orbit would represent a fine-tuning problem in this work, given our choice of model that neglects poorly-constrained physical effects such as radius inflation (see \autoref{appendix} for further details). Here we provide a brief outline of how the methods of this work differ from those of \Atwentyfive.

\begin{enumerate}[label=(\alph*)]
    \item \textit{Dynamical Methods}: The most major difference is the equations of motion themselves. Both studies integrate the orbit-averaged equations of \cite{eggleton_2001}, appropriate for hierarchical triple systems. However, \Atwentyfive truncates these equations to second order in $a_\mathrm{b} / a_\mathrm{c}$, the quadrupole approximation. In contrast, our study uses the publicly available \texttt{kozaipy} package, which supports integration of the equations of motion to third order in $a_\mathrm{b} / a_\mathrm{c}$, the octupole approximation. 

    Numerous works \citep[e.g.][]{naoz2013, li_eccentricity_2014} have noted that considering this additional order in the equations of motion results in qualitatively different behavior. For our purposes, the most significant of these is secular descent \citep{weldon2024descent}. If octupole-order effects are accounted for, the amplitude of ZLK-driven eccentricity oscillations does not remain constant. Rather, the eccentricity maxima will slowly grow to extreme values \citep{teyssandier2013orbitflips,Li14,li_eccentricity_2014}. This is not present in the simulations of \Atwentyfive, who report a maximum eccentricity over all simulations of $e\sim0.94$. Our simulations, which account for octupole-order effects, reach higher eccentricities. These differences lead to qualitatively distinct dynamical evolution: the rates of both tidal dissipation and ZLK precession scale as $ \propto e^2$ \citep{millholland2019obliquity}, and the rate of tidal precession scales as $\propto e^{-6}$ \citep{fabrycky2007shrinking}. In \autoref{fig:epsilon_oct}, we show the distribution of the octupole coefficient $\epsilon_{\mathrm{Oct}} = \frac{a_{\mathrm{b}}}{a_{\mathrm{c}}} \frac{e_{\mathrm{c}}}{1 - e_{\mathrm{c}}^2}$, which characterizes the strength of octupole-order perturbations. Given that a substantial fraction of the configurations achieve $\epsilon_{\mathrm{Oct}}>0.03$ (shown to be the nominal limit at which octupole-order effects become very significant for $I_{\mathrm{mut,i}}>60^\circ$ by \citealt{MunozLai2016}), we expect many of our simulations to reach eccentricities beyond what would be predicted in the quadrupole approximation.\footnote{The octupole-order maximum eccentricity is highly sensitive to the initial eccentricity and often varies significantly in comparison to the quadrupole maximum eccentricity (see \citealt{LithwickNaoz2011,teyssandier2013orbitflips,li_eccentricity_2014,weldon2025cold}).}
    
    In addition, our tidal models differ. \Atwentyfive uses a constant $Q$ model, and we use a constant time lag model. Constant $Q$ models assume tidal misalignments independent of forcing frequency---an assumption based on laboratory experiments of energy dissipation in rocky materials \citep{knopoff_64, goldreich1966} that is appropriate for fully solid bodies, such as rocky planets lacking a substantial gaseous envelope. The constant time lag model is derived in the visco-elastic limit, and as such is more appropriate for gaseous planets \citep{zahn1977tidal, ogilvie2004tidal}.
    
    \item \textit{Observational constraints}: \Atwentyfive provide observational constraints using the HARPS subset (4.19 years) of RV measurements analyzed in this work. We incorporate 15 more years of baseline archival RV data, along with astrometric constraints from the HGCA.
    
    \item \textit{Photoevaporation modeling}: \Atwentyfive uses the \texttt{JADE} code \citep{attia2021JADE}, which incorporates detailed modeling of photoevaporation-driven mass loss induced by strong ultraviolet (XUV) radiation. The effect of mass loss is not considered within our analysis. However, \Atwentyfive find that GJ 436 b has experienced minimal cumulative mass loss throughout its evolution. Thus, while our models do not consider the impact of mass loss, the results of \Atwentyfive suggest that this effect should not appreciably alter the dynamics of the system.  

\end{enumerate}

\section{Conclusions  \label{Conclusions}}

The eccentric and polar orbit of GJ 436 b is a dynamical mystery that can be reproduced through ZLK migration, which would imply the presence of an unseen external perturber. In this study, we present the most comprehensive dynamical characterization of the GJ 436 system to date, placing stringent limits on the mass and orbital properties of a putative outer perturber. 

Our findings are summarized as follows:
\begin{enumerate}[label=(\alph*)]

\item By conducting a joint analysis of $\sim$20 years of archival RV measurements from three independent spectrographs (Keck/HIRES, HARPS, and CARMENES) together with absolute astrometry from the HGCA catalog, we exclude perturbers with $a_{\mathrm{c}}<5.4$~AU for $m_{\mathrm{c}}>0.05~M_{\mathrm{Jup}}$ and $a_{\mathrm{c}}<64$~AU for $m_{\mathrm{c}}>24~ M_{\mathrm{Jup}}$ in the GJ 436 system at the $2\sigma$ confidence level.
\item Through dynamical modeling of 1,000 companion configurations drawn from our observational posteriors across 18 initial condition suites, we demonstrate that ZLK migration can successfully reproduce GJ 436 b's orbit, including both the residual eccentricity and the polar orbit. 
\item Our dynamical models reveal that the GJ 436 b orbital configuration can be well-reproduced by interactions with a sub-Jovian planetary companion spanning $m_{\rm c}=0.55_{-0.39}^{+1.67}~M_{\text{Jup}}$ (16th--84th percentile) on a wide orbit of $a_\mathrm{c} \gtrsim 6.8$ AU.
\end{enumerate}

It is worth noting that our study did not account for the effects of planetary structural evolution over the course of orbital migration, which has been the subject of great interest in recent years \citep[e.g.][]{attia2021JADE,yu_dai_2024,LuAn2025,hallattmillholland2025desert,hallattmillholland2025coupled,attia2025JADE}. While \Atwentyfive indicated that mass loss driven by photoevaporation in GJ 436 b---for which there is observational evidence \citep{LavieEhrenreich2017}---is likely not dynamically significant, radius inflation driven by tidal heating \citep[e.g.][]{millholland_2019,millholland2020tidal, LuAn2025, SethiMillholland2025} and Roche lobe overflow \citep[e.g.][]{valsecchi2014hot, valsecchi2015roche,weldon2025saving} may play a role in this system.

The GJ 436 system serves as a valuable archetype for understanding the formation and evolution of polar hot Neptunes in the sub-Jovian desert. Based on our predictions, the detection of a potential outer companion in the GJ 436 system would likely add to the growing census of multi-planet systems observed with large mutual inclinations between orbits \citep{McArthurBenedict2010, 2011ApJ_Sanchis-Ojeda_hatp11_star_spin, MillsFabrycky2017, XuanWyatt2020, BourrierLovis2021,An2025significant, BardalezGagliuffiBalmer2025, ZhangWeiss2025}.
These systems highlight the significant role that planet-planet interactions play in sculpting planetary system architectures. While the upcoming Gaia Data Release 4 is unlikely to detect GJ 436c if its properties match those predicted in this work \citep{lammers2025exoplanet}, future high-precision direct imaging surveys may prove more fruitful.

\section*{acknowledgements}
We thank the anonymous referee for comments that greatly strengthened the manuscript. We thank Greg Laughlin for early conversations that inspired this work, and Diego Mu\~{n}oz, Shangjia Zhang, Helen Qu and the Astronomical Data Group at the CCA for insightful conversations. We are grateful to the Dorrit Hoffleit Undergraduate Research Scholarship program at Yale University, which provided support for this project. T.L. is supported by a Flatiron Research
Fellowship at the Flatiron Institute, a division of the Simons Foundation. M.R. acknowledges support from Heising-Simons Foundation Grants \#2021-2802 and \#2023-4478, National Geographic Grant EC-115062R-24, and the NASA Exoplanets Research Program NNH23ZDA001N-XRP (grant
\#80NSSC24K0153). This work has benefited from use of the \texttt{GRACE} computing cluster at the Yale Center for Research Computing (YCRC).

\appendix 
\section{Radius Inflation}
\label{appendix}
\begin{figure}[h]
    \centering
    \includegraphics[width=0.5\linewidth]{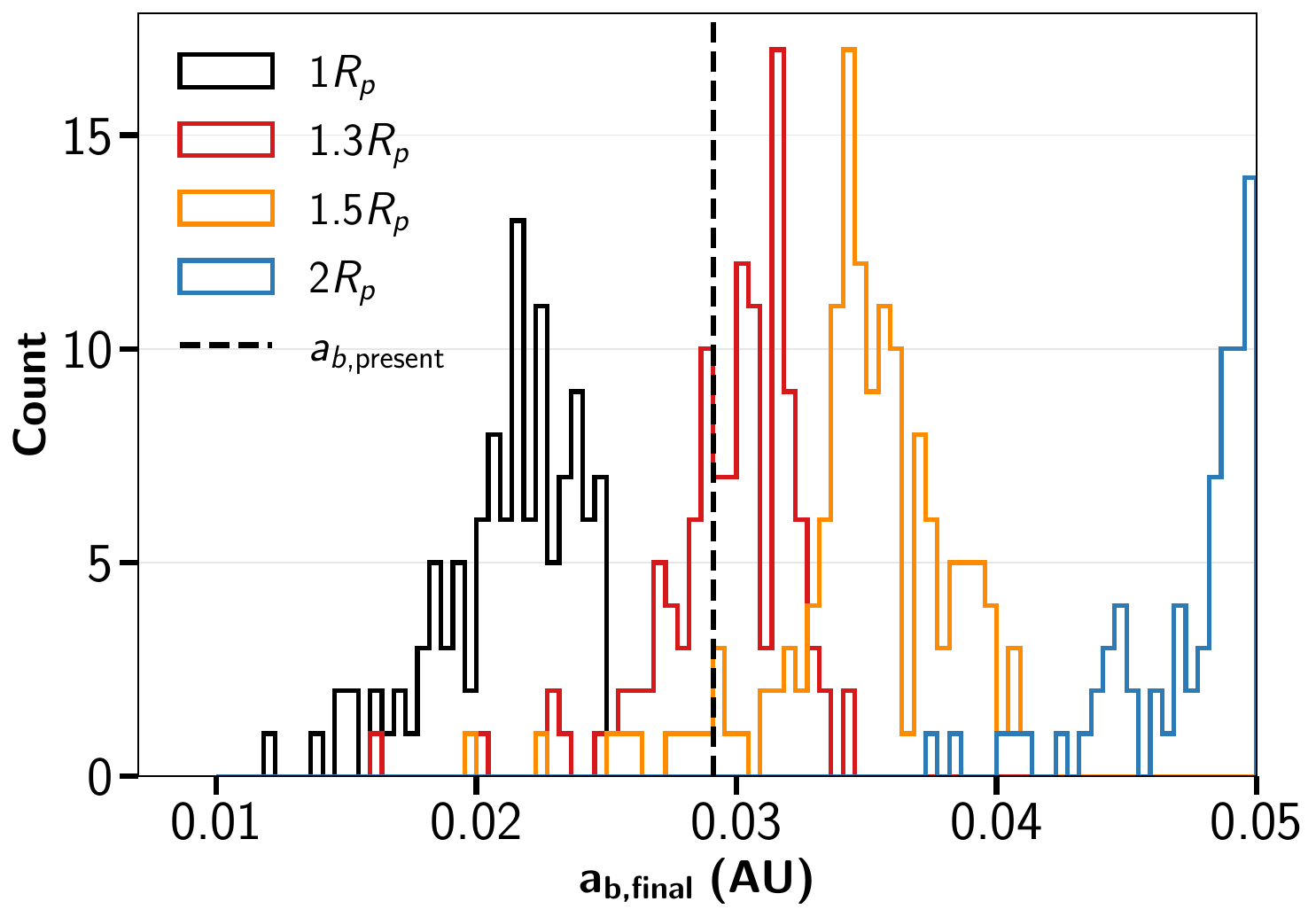}
    \caption{Distribution of final semi-major axes $a_{\mathrm{b,final}}$ for plausible simulations initialized with $I_{\mathrm{mut,i}}=80^\circ$ and $a_{\mathrm{b,i}}=1$ AU across four planetary radius factors: 1$R_{\mathrm{p}}$ (black), 1.3$R_{\mathrm{p}}$ (red), 1.5 $R_{\mathrm{p}}$ (orange), and 2$R_{\mathrm{p}}$ (blue). The dashed vertical line marks the present-day observed semi-major axis of GJ 436 b (0.0291 AU).}
    \label{fig:finalaplot}
\end{figure}

The median $a_{\mathrm{b,final}}$ across our plausible simulations ranges from 0.016 to 0.022 AU depending on the initial semi-major axis, roughly a factor of ${\sim}1.3-1.8 \times$ smaller than the final semi-major axis predicted by present-day observations. This discrepancy can be resolved by invoking tidally-driven radius inflation \citep{BodenheimerLin2001, millholland_2019, millholland2020tidal}. \cite{LuAn2025} demonstrated that if radius inflation is considered self-consistently with system dynamics, $a_{\text{b,final}}$ in ZLK migration may be increased by up to a factor of ${\sim}2$.

While a fully self-consistent dynamical model incorporating radius inflation is beyond the scope of this work, we offer a simple demonstration of the significant effect radius inflation may play. We have run three additional suites of 1,000 companion configurations each, with initial conditions corresponding to the $a_{\mathrm{b,i}} =1$~AU and $I_{\mathrm{mut,i}} = 80^\circ$ simulations described in \S\ref{simulation_setup}. In these three additional suites, we inflate the planetary radius by $1.3\times$, $1.5\times$, and $2.0\times$ the present-day $R_{\mathrm{p}}$. \autoref{fig:finalaplot} shows the resulting distributions of $a_{\mathrm{b,final}}$ for simulations classified as plausible along with the $1R_{\mathrm{p}}$ distribution. As expected, larger radii shift $a_{\mathrm{b,final}}$ towards larger values with the $R = 1.3 R_\mathrm{p}$ bin consistent with the present-day semi-major axis.

This is a smaller effect than predicted by \cite{LuAn2025}. We emphasize that the results of \cite{LuAn2025} are highly model-dependent. The true degree of radius inflation is not well-constrained given the considerable observational uncertainties in quantities such as tidal quality factor; however, the projected range of possibilities easily encompasses what is necessary to generate GJ 436b's present-day semimajor axis.

\software{\texttt{emcee} \citep{Foreman-MackeyFarr2019}, \texttt{ethraid} \citep{VanZandtPetigura2024}, \texttt{kozaipy}, \texttt{RadVel} \citep{FultonPetigura2018}, \texttt{orvara} \citep{BrandtDupuy2021}, \texttt{matplotlib} \citep{hunter_2007},  \texttt{numpy} \citep{oliphantnumpy, vanderwalt2011numpy,harris2020array} }

\bibliography{sample7}{}

\end{CJK*}
\end{document}